\documentclass[reprint, aps,prb, amsmath, amssymb,longbibliography]{revtex4-2}
\usepackage[dvipdfmx]{graphicx}
\usepackage{color}
\usepackage{dcolumn}
\usepackage{bm}
\usepackage{here}
\usepackage{placeins}
\usepackage{comment}
\usepackage{pdfpages} 
\usepackage{pgffor} 
\makeatletter
\AtBeginDocument{\let\LS@rot\@undefined}
\makeatother

\def\supplementfilename{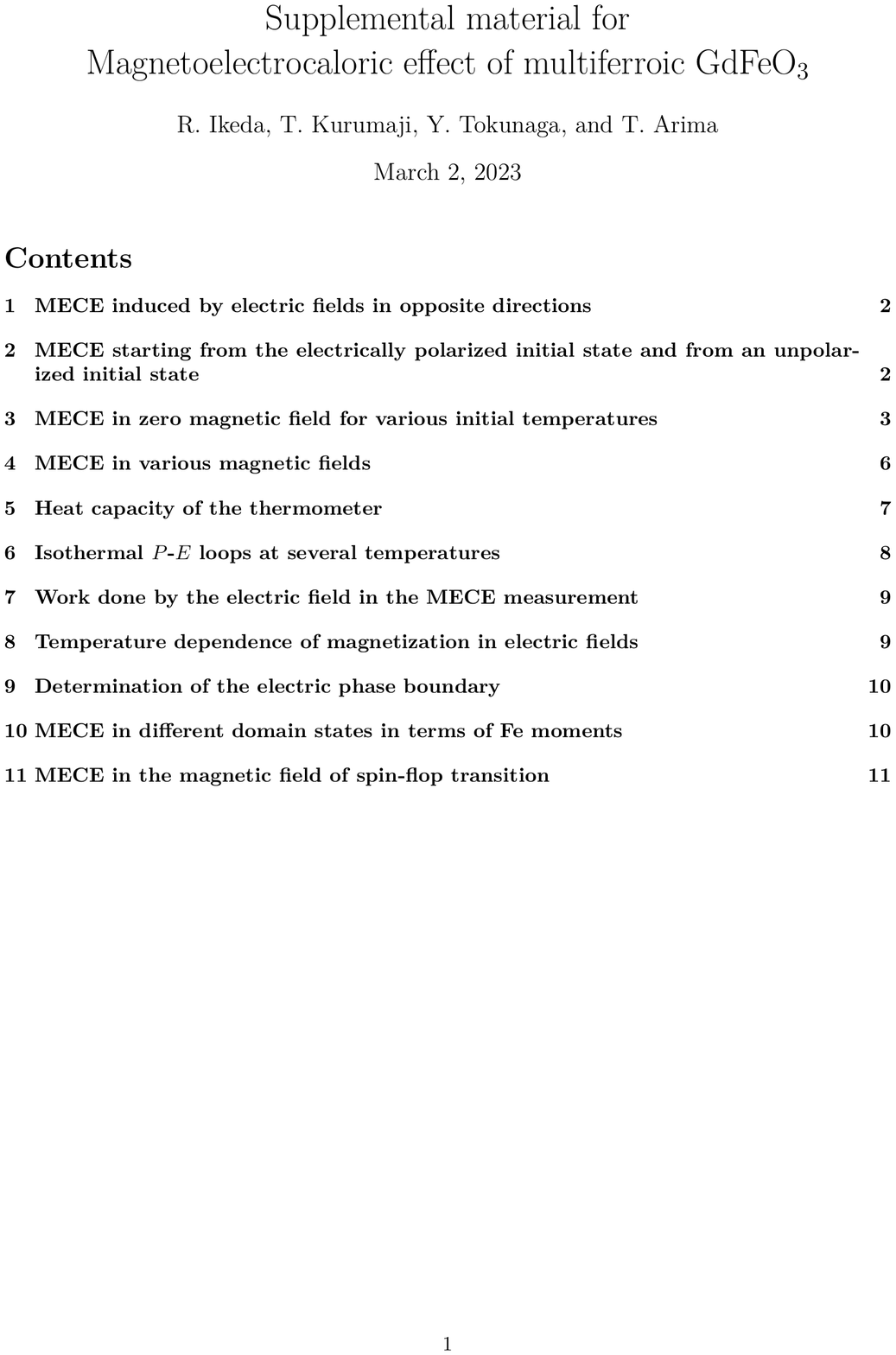}

\pdfximage{\supplementfilename}
\def\numbersupplementpages{\the\pdflastximagepages}

\newif\ifarXiv
\arXivtrue 

\begin{document}


\title{Magnetoelectrocaloric effect of multiferroic $\mathrm{GdFeO_3}$}


\author{Rintaro Ikeda, Takashi Kurumaji, Yusuke Tokunaga, and Taka-hisa Arima}
\affiliation{Department of Advanced Materials Science, The University of Tokyo, Kashiwa 277-8561, Japan}


\date{\today}

\begin{abstract}
We report an experimental demonstration of the magnetoelectrocaloric effect (MECE) in multiferroic $\mathrm{GdFeO_3}$. The temperature of the magnetic material changes when an external electric field is suddenly changed. MECE is the largest below the ordering temperature of Gd moments and modifiable by a magnetic field, suggesting that the ferroelectric transition affects MECE. The observed MECE shows a higher energy efficiency than typical magnetocaloric effects and adiabatic nuclear demagnetization. The present findings provide a proof of concept of MECE in multiferroics.
\end{abstract}

\maketitle

Various cooling methods based on magnetocaloric effect (MCE) and adiabatic nuclear demagnetization (AND) have been investigated due to increasing demand for highly energy-efficient cooling devices applicable to hydrogen liquefaction and quantum computers. While MCE and AND, exploiting entropy changes with the magnetic-field application and removal, are used in laboratories, the current-driven magnetic cooling methods have some disadvantages: the electric current to apply the magnetic field consumes energy; superconducting magnets are bulky; and shielding of the magnetic field is required in some cases.  The magnetic refrigeration using permanent magnets needs additional maintenance for the mobile system and causes mechanical vibration and noise.

\begin{figure}[h]
\includegraphics{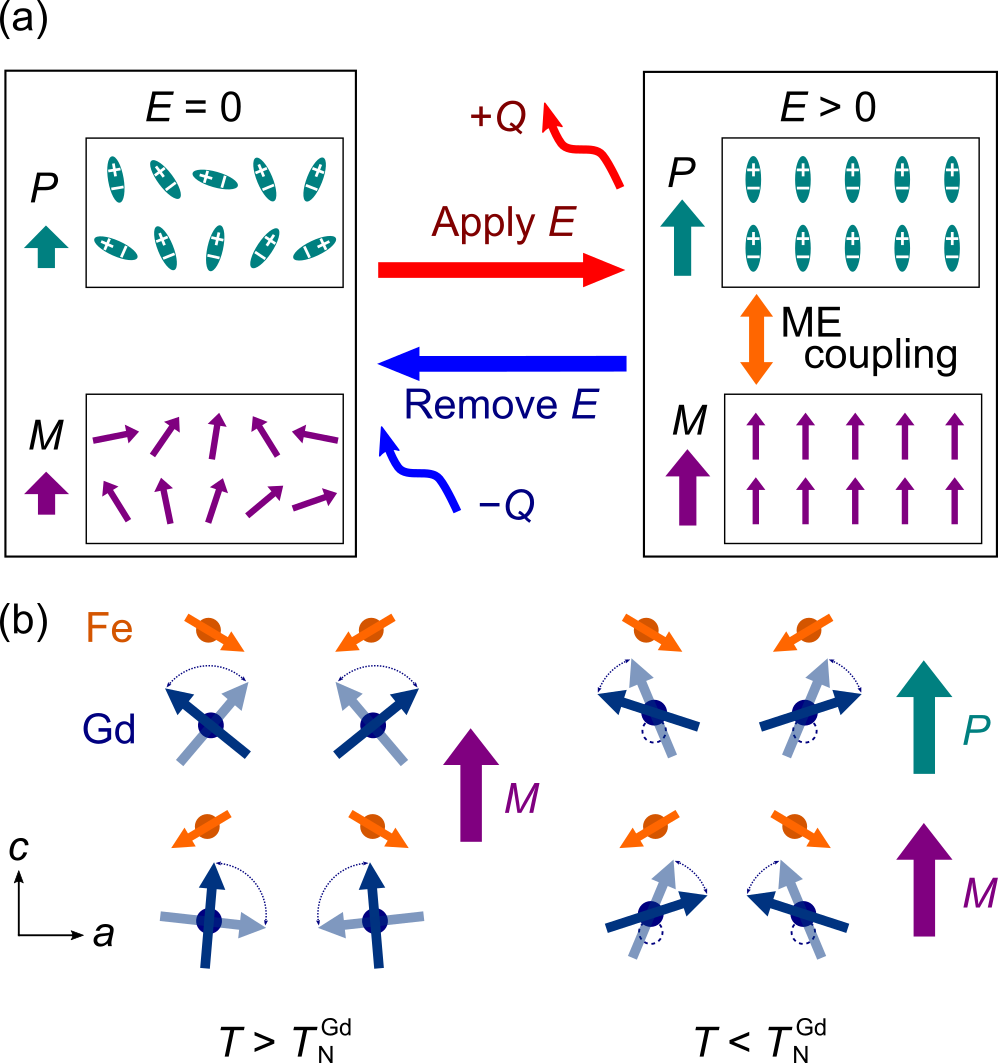}
\caption{(a) Schematic diagram of magnetoelectrocaloric effect (MECE) in multiferroics. Exothermic (endothermic) effect, release (absorption) of heat $Q$ to (from) the lattice system, happens by the entropy change with an application (removal) of an electric field $E$ in the process from left (right) to right (left). Electric polarization $P$ increases (decreases) by aligning (misaligning) electric dipolar moments (green ellipses), which accompanies the increment (reduction) of magnetization $M$ by ordering (disordering) of magnetic moments (purple arrows) through magnetoelectric (ME) coupling. 
(b) Schematic magnetic structures of $\mathrm{GdFeO_3}$ above and below the transition temperature $T_{\mathrm{N}}^{\mathrm{Gd}}$ of Gd moments. Fe spins (orange arrows) are antiferromagnetically arranged with canting to host $M$ parallel to the $c$ axis. Dark and light blue arrows represent Gd spins. Gd spins are fluctuating around the direction of the exchange magnetic field produced by Fe spins. Below $T_{\mathrm{N}}^{\mathrm{Gd}}$, Gd atoms (blue circles) are displaced from their original positions (dotted circles) to induce $P$ along the $c$ axis.
}
\label{fig_scheme}
\end{figure}

Recently, the multicaloric effect, more than one type of caloric effects simultaneously driven by a single external field (magnetic, electric, or stress), has attracted much attention \cite{Starkov2014, APL_MultiferroicComposites, Moya2014}. A multicaloric effect termed magnetoelectrocaloric effect (MECE) is expected not only to enhance MCE and electrocaloric effect (ECE) but also to achieve high tunability since other external fields modify the phase transition temperature to optimize the caloric effect \cite{Moya2014}.

The MECE is anticipated in multiferroic materials, where the magnetism and electricity are strongly correlated. A schematic illustration of MECE in a multiferroic material with a $P^2M^2$ term in the free energy is shown in Fig. \ref{fig_scheme}(a). At zero field (left panel in Fig. \ref{fig_scheme}(a)), the system is in a heavily fluctuating state. When an electric field is applied, the entropy of the electric dipole moments decreases since they are aligned. The magnetic entropy is also reduced through the ME coupling (right panel in Fig. \ref{fig_scheme}(a)). The reduced entropy is transferred to the lattice system, resulting in a temperature rise in the adiabatic condition. With the removal of the electric field, the electric and magnetic dipole moments must regain the entropy from the lattice system.

Due to the strong ME coupling, MECE provides an opportunity to control the entropy in both electric and magnetic systems with an electric field. As no electric current is  necessary, the energy dissipation due to Joule heating is expected to be smaller than that in MCE and AND. Despite some theoretical predictions \cite{PRLmC, PhysRevB.98.174105, VOPSON20122067, doi:10.1063/PT.3.3022, Stern-Taulats2018}, experimental direct measurements of the MECE have been scarcely reported. In this paper, we report an observation of MECE in a multiferroic $\mathrm{GdFeO_3}$.

$\mathrm{GdFeO_3}$ is known to possess the controllability of magnetism and ferroelectric polarization by electric fields $E$ and magnetic fields $H$, respectively \cite{Tokunaga2009}. Fe spins ($S_{\mathrm{Fe}} = 5/2$) are antiferromagnetically arranged below $T_{\mathrm{N}}^{\mathrm{Fe}}$ = 661 K \cite{doi:10.1063/1.1714088} with tiny canting to induce weak magnetic moment $M$ along the $c$ axis in the $Pbnm$ setting (see the left panel of Fig. \ref{fig_scheme}(b)) \cite{DZYALOSHINSKY1958241, PhysRev.120.91}. Gd spins ($S_{\mathrm{Gd}} = 7/2$) with negligible anisotropy are antiferromagnetically arranged at $T_{\mathrm{N}}^{\mathrm{Gd}}$ = 2.5 K \cite{ges198757}. Canting of Gd spins contributes to $M$, which is antiparallel to that induced by canted Fe spins. Below $T_{\mathrm{N}}^{\mathrm{Gd}}$, symmetric exchange striction due to Gd-Fe interaction modifies the distance between these ions to generate polarization $P$ along the $c$ axis. The ferroelectric transition hence coincides with the antiferromagnetic transition of Gd spins. Most of the magnetic entropy $R\ln8$ of Gd spins is released around $T_{\mathrm{N}}^{\mathrm{Gd}}$ \cite{PhysRevB.96.174405}. These features make $\mathrm{GdFeO_3}$ a promising candidate for the observation of MECE.

A single crystal of $\mathrm{GdFeO_3}$ was grown by the floating zone method in an oxygen atmosphere. The phase purity was confirmed by powder X-ray diffraction. The crystallographic orientation was determined by X-ray Laue photographs. The crystal rod was cut and polished to obtain smooth faces of (001) for the parallel-plate capacitor structure. The dimensions of the processed sample for measurements of MECE ($P$ and $M$) were approximately $2\times3\times0.1$ ($2\times1\times0.035$) $\mathrm{mm^3}$. Heat-treatment-type silver paste was deposited on both the large surfaces of the sample to form electrodes for the measurement of $P$ and $M$. To measure $P$-$E$ hysteresis loops and temperature dependence of $M$ in $E$, $E$ and $H$ fields are applied along the $c$ and $a$ axes, respectively. $M$ was measured by a superconducting quantum interference device magnetometer (MPMS, Quantum Design) with a measurement probe capable of the application of an electric field with an electrometer (model 6517A, Keithley). $P$ was simultaneously measured using the Q mode of the electrometer. Heat capacity measurements were performed by using heat capacity option of a super conducting magnet PPMS, Quantum Design. The peak temperature of the specific heat in zero magnetic field suggested that $T_{\mathrm{N}}^{\mathrm{Gd}}$ of the crystal was 2.25 K.

\begin{figure}[t]
\includegraphics{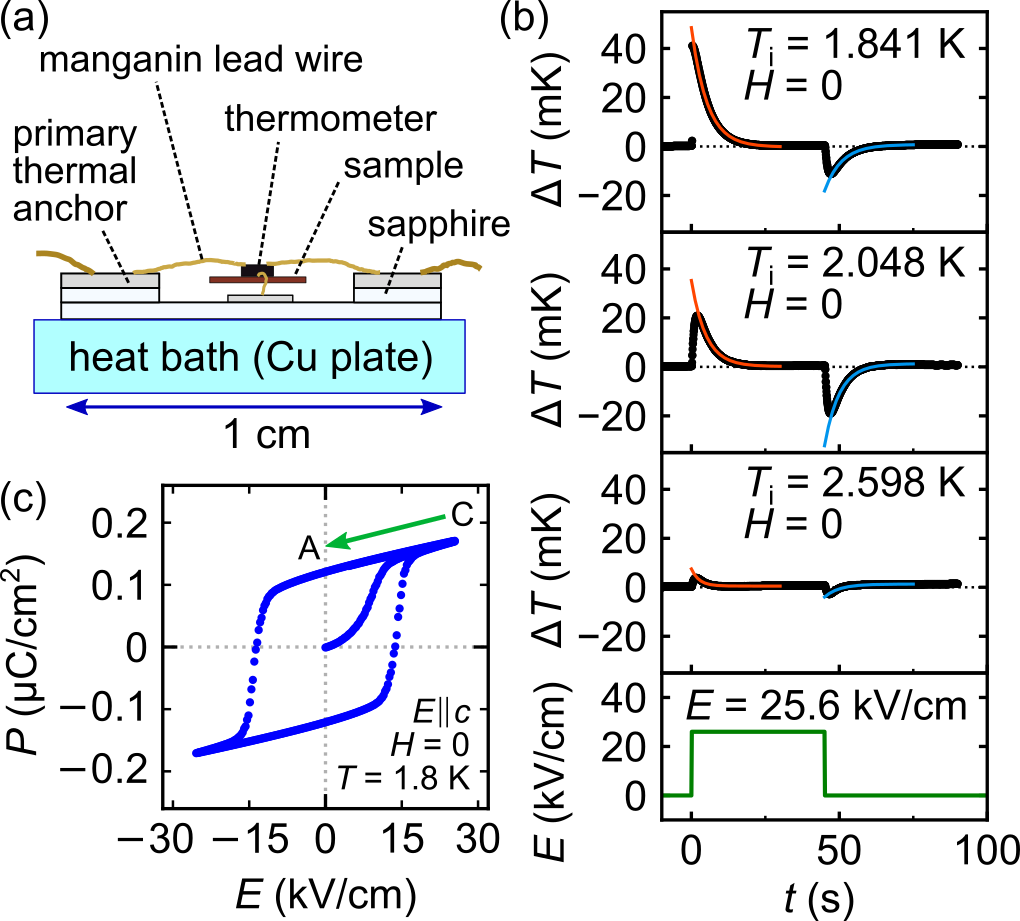}%
\caption{(a) Side view of the setup for MECE measurements. A thin plate sample is maintained in the quasi-adiabatic condition by being hung in vacuum with manganin wires from primary thermal anchors made of sapphire (grey), which are mounted on a copper heat bath. Primary thermal anchors are connected to secondary thermal anchors with copper wires (yellow thick lines).
(b) Time profiles of the applied electric field $E$ and the change of the sample temperature $\Delta T$ for several initial sample temperatures $T_\mathrm{i}$ in zero field. Red and blue curves are the simple exponential decay (see text).
(c) Isothermal $P$-$E$ hysteresis loop at 1.8 K. The external electric field $E$ is applied along the $c$ axis. The green arrow is the isothermal $E$-removal process, corresponding to path C-A in Fig. \ref{fig_TEgraph}(a). 
}
\label{fig_tempvstime}
\end{figure}

For a direct measurement of MECE of $\mathrm{GdFeO_3}$ in the adiabatic condition, we construct an experimental setup, as shown in Fig. \ref{fig_tempvstime}(a), which was attached to the probe to load into a PPMS. The sample temperature $T_{\mathrm{s}}(t)$ was monitored every 50 ms by a temperature controller (model 340, Lake Shore Cryotronics) and a calibrated resistance thermometer (CX-1030-BR-HT, Lake Shore Cryotronics), which was thermally contacted to the sample by grease. Heat-bath temperature $T_{\mathrm{b}}$ was also monitored with another thermometer on the heat bath. Lead wires made of manganin provided structural support for hanging the sample and the thermometer in vacuum. Gold films of 3000 $\text{\AA}$ thickness were sputtered on the both sides of the sample to apply an electric field parallel to the $c$ axis. The chamber atmosphere was evacuated to 1 Pa with a cryogenic pump in order to maintain the adiabatic condition of the sample. The relaxation of the sample temperature was dominated by the thermal conduction through the manganin wires. A magnetic field was applied along the $a$ axis. Each MECE measurement started after the sample was cooled from a temperature above $T_{\mathrm{N}}^{\mathrm{Gd}}$ with $E$ = 0.

\begin{figure}[h]
\begin{center}
\includegraphics{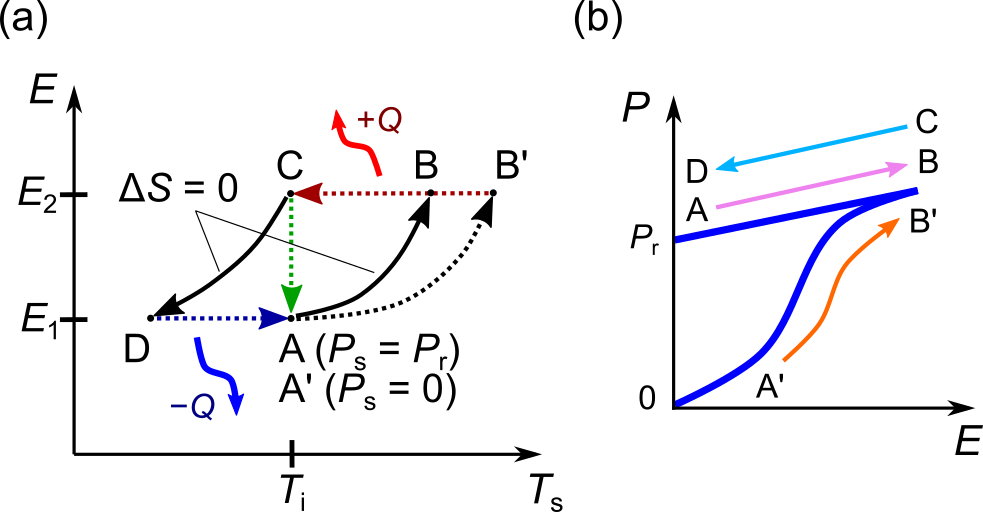}%
\caption{(a) Schematic drawing of the trajectory of the MECE measurement in the $T$-$E$ plane. $T_{\mathrm{i}}$ is the initial temperature of the sample, which is equilibrated with the heat bath. Solid (dashed) lines represent an isentropic (non-isentropic) processes. $\Delta S_{\mathrm{direct}}$ in Eq. (\ref{eq_Sdirect}) is calculated in the warming (cooling) process represented by the blue (red) line. $\Delta S_{\mathrm{indirect}}$ in Eq. (\ref{eq_Sindirect}) is calculated in the isothermal process represented by the green line. A (A') represents the initial state with the non-zero (zero) residual polarization $P_{\mathrm{r}}$. The state of magnetic and electric dipole moments in A (C) corresponds to the left (right) panel of Fig. \ref{fig_scheme}(a).
(b) Schematic adiabatic $P$-$E$ curve. The light red (orange) arrow represents the $E$-application process, path A-B (A'-B'). 
}
\label{fig_TEgraph}
\end{center}
\end{figure}

\begin{figure}[h]
\includegraphics[width=8.1cm]{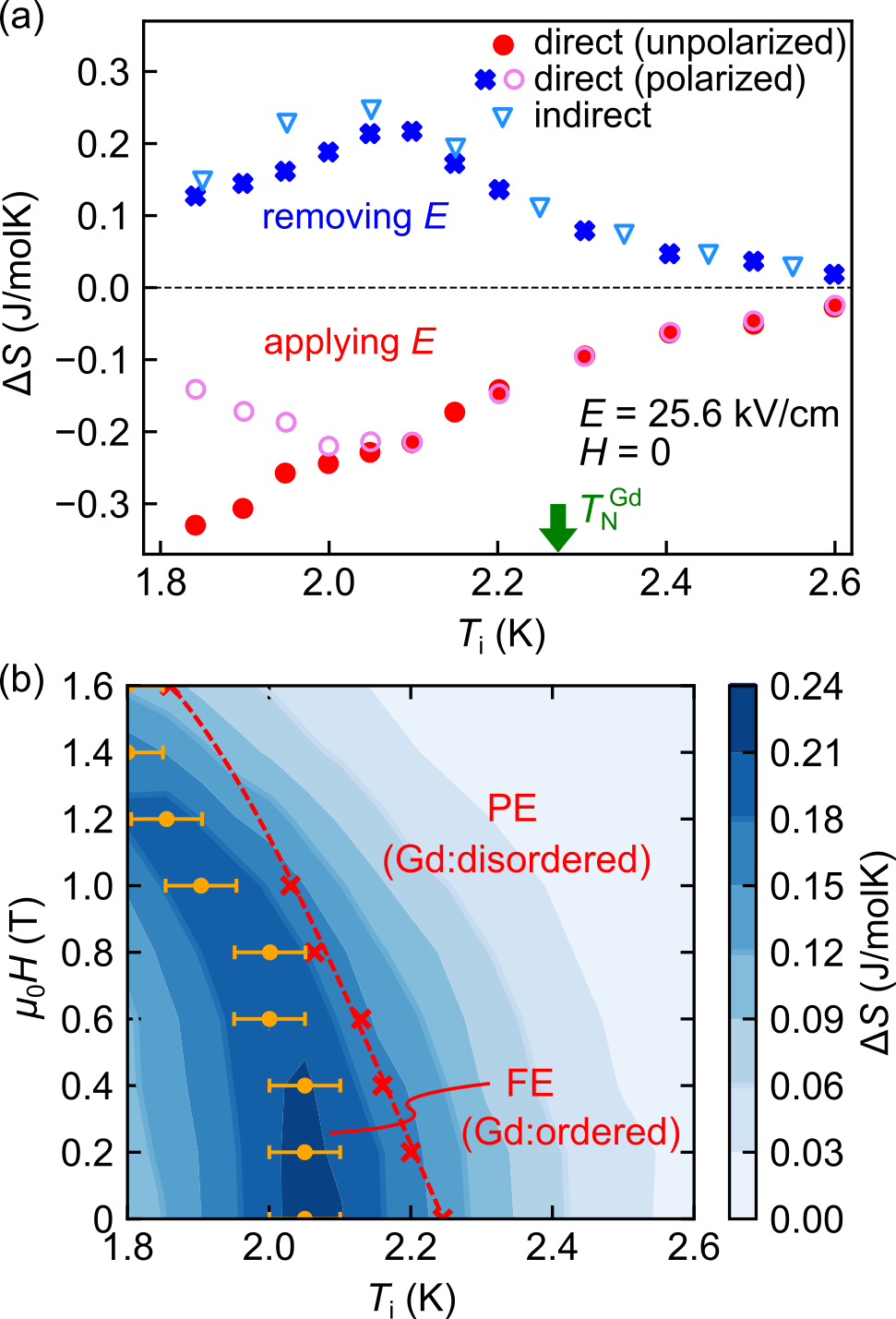}%
\caption{(a) Initial sample temperature $T_{\mathrm{i}}$ dependence of entropy change of  $\Delta S$, estimated with Eq. (\ref{eq_Sdirect}), when an electric field of 25.6 kV/cm is applied (red circles) and removed (blue crosses) in zero field. Light red open circles are $\Delta S$ in the application process when the sample had been polarized before the measurement. The polarization changes of the application and removal processes represented by red circles, pink open circles, and blue crosses are shown with the red (path A'-B'), light red (path A-B), and blue (path D-A) arrows, respectively, in Fig. \ref{fig_TEgraph}(b). Cyan triangles represent the indirect estimation of $\Delta S$ calculated from Eq. (\ref{eq_Sindirect}). The transition temperature for Gd moments $T_{\mathrm{N}}^{\mathrm{Gd}}$ determined by a heat capacity measurement is denoted by the green arrow. Each temperature change of the sample is shown in the supplemental material. 
(b) Color map of the entropy change $\Delta S$ by the removal of an electric-field of 25.6 kV/cm in the $H$-$T$ phase diagram for $H \parallel a$. Red crosses are the peak of specific heat (SI Fig. S9) at several static magnetic fields. Dashed line is a guide to the eye. PE (FE) denotes the paraelectric (ferroelectric) phase with disordered (ordered) Gd moments. Orange circles represent the temperature at which the entropy increase is maximized for each magnetic field.
}
\label{fig_phasediagram}
\end{figure}

In Fig. \ref{fig_tempvstime}(b), we present time $t$ dependence of temperature change $\Delta T(t) = T_{\mathrm{s}}(t) - T_{\mathrm{s}}(0)$ of the sample at zero field for various initial temperatures $T_{\mathrm{i}}$ ($= T_{\mathrm{s}}(0)$). We observe a sharp exothermic (endothermic) peak upon the application (removal) of $E$, which is followed by an exponential relaxation to the initial temperature. The relaxation excludes the Joule heating effect. We confirm the identical temperature changes in the opposite electric field \cite{suppl}, which is consistent with the mechanism that the MECE is independent of the direction of $E$. A negligible temperature change is observed for $T_{\mathrm{i}}$ = 2.6 $\mathrm{K}>T_{\mathrm{N}}^{\mathrm{Gd}}$, indicating the Gd-spin origin of this MECE effect.

We note that the temperature increase in the exothermal process is larger than the magnitude of the temperature decrease in the endothermal process (see the top panel of Fig. \ref{fig_tempvstime}(b), for example). This partly stems from the dissipation by the domain wall motion in the $E$-application process.

To capture the $E$-driven evolution of $P$, we show an isothermal $P$-$E$ loop at 1.8 K in Fig. \ref{fig_tempvstime}(c). The maximum $E$ of the $P$-$E$ hysteresis is the same as $E$ applied in the MECE measurement, which surpasses the coercive electric field (\textasciitilde 13 kV/cm) to fully polarize the ferroelectric domain. The fully polarized state is maintained at $E$ = 0 after the removal of $E$ below $T_{\mathrm{N}}^{\mathrm{Gd}}$ (green arrow in Fig. 2(c)).

The heating and cooling of the sample through MECE can be described in the $E$-$T$ diagram (Fig. \ref{fig_TEgraph}(a)). Corresponding $P$-$E$ curve in the adiabatic condition is also shown in Fig. \ref{fig_TEgraph}(b). Arrows represent the $E$-change process and following thermal relaxation process. In the ideal isentropic process, where the ferroelectric monodomain state is maintained, the temperature increase with the application of $E$ can be represented by path A-B. Path B-C represents the temperature relaxation to the heat bath temperature due to thermal conductance of manganin wires. C-D and D-A represent in the $E$-removal process and the following temperature relaxation, respectively. When the initial state is not polarized due to multidomain cancellation of $P$ (state A' in Fig. \ref{fig_TEgraph}(a)), the additional heating in the $E$-application process warms up the system to state B'.

The entropy change $\Delta S_{\mathrm{direct}}$ through MECE can be estimated directly from the temperature changes $\Delta T_{E}$ on path B-C (B'-C) and D-A as 
\begin{equation}
    \Delta S_{\rm{direct}} = \frac{C(T_{\mathrm{i}}) \Delta T_E}{T_{\mathrm{i}}},
    \label{eq_Sdirect}
\end{equation}
where $C(T_{\mathrm{i}})$ is the total heat capacity of the sample and thermometer at $T_{\mathrm{i}}$. To obtain the temperature change $\Delta T_{\mathrm{E}}$ at states B, B', and D, we extrapolate the temporal change $\Delta T(t)$ by an exponential function to the moment of the $E$-change, as shown in Fig. \ref{fig_tempvstime}(b). In our experimental setup, the heat capacity of the thermometer is negligibly small compared to that of the sample \cite{suppl}. We also estimate the entropy change $\Delta S_{\mathrm{indirect}}$ on the isothermal path C-A (green line in Fig. \ref{fig_TEgraph}(a)) by using Maxwell's relations as 
\begin{equation}
    \Delta S_{\rm{indirect}} = -V\int_{E_{i}}^{E_{f}}\left(\frac{\partial P}{\partial T}\right)_{E} dE,
    \label{eq_Sindirect}
\end{equation}
where $E_i$ ($E_f$) is the initial (final) value of the applied electric field, and $V$ is the volume. $\frac{\partial P}{\partial T}$ is calculated from the isothermal $P$-$E$ loop at various temperatures (see Supplement).

Figure \ref{fig_phasediagram}(a) summarizes the estimated entropy changes. Significant entropy changes are observed below $T_{\mathrm{N}}^{\mathrm{Gd}}$, suggesting that the entropy change originates from the ordering of Gd spins. For the endothermal processes, both estimations show a good agreement with each other, indicating that the isentropic condition (path D-A) is well satisfied. Temperature dependence of the decrease in entropy with the application of an electric field along the light red arrow in Fig. \ref{fig_TEgraph}(b) after the sample is polarized with $E$ = 25.6 kV/cm (light red open circles in Fig. \ref{fig_phasediagram}(a)), is nearly consistent with the increase in entropy with the removal of $E$ (blue crosses), which also suggests that the entropy change arises from MECE. This is because both paths A-B and C-D go through the same path in the adiabatic $P$-$E$ hysteresis (Fig. \ref{fig_TEgraph}(b)).

$\Delta S$ in the $E$-removal process becomes the largest at $T_{\mathrm{i}}$ = 2.05 K. The reduction of $\Delta$S below the temperature is consistent with a prediction that $\Delta S$ should be zero at the absolute zero temperature because spins have no kinetic energy. It presumably explains the deviation of the entropy decrease (red circle in Fig. \ref{fig_phasediagram}(a)) in the $E$-application at the lowest temperature, $T$ = 1.85 K.

There is a gap between $\Delta S$ on path A-B and $\Delta S$ on path A'-B' in the $E$-application process in Fig. \ref{fig_phasediagram}(a). The additional heating when starting from the unpolarized state is owing to the electric work for domain reversal \cite{suppl}. The discrepancy between the two cases appears below 2 K and increases as the temperature is lowered, which qualitatively agrees with the estimated work by $E$. Besides, the entropy change is independent of the domain state of Fe spins \cite{suppl}. This result shows that it is enough to discuss the ferroelectric $p$ domain and magnetic domain of Gd $f$ moments since the $p$ domain and the $f$ domain simultaneously change under the fixed phase (0 or $\pi$) of staggered alignment of Fe moments \cite{Tokunaga2009}. In the context of MCE (ECE), domain rotation and magnetic (electric) domain walls are reported to contribute to heating and cooling \cite{DW_movement_rotation_MCE, DW_MEC_theoretical, DW_MCE_experimental, DW_movement_ECE, DW_ECE_1, DW_ECE_2, DW_ECE_3, DW_ECE_4}.

We observe a large MECE in the ferroelectric (FE) phase, where Gd spins are ordered, in the $H$-$T$ phase diagram (Fig. \ref{fig_phasediagram}(b)). MECE is reduced in the paraelectric (PE) phase. The temperature where MECE is maximized decreases as a larger magnetic field is applied, in accord with the reduction of the FE transition temperature determined by the specific heat measurement \cite{suppl}. We do not observe any apparent anomalies associated with the spin flop transition where both Gd and Fe spins rotate by 90 degrees at 0.5 T \cite{Tokunaga2009} (see SI Fig. S12).

Temperature dependence of magnetization in electric fields (SI Fig. S8) suggests that an electric field enhances antiferromagnetic ordering of Gd moments only below the ferroelectric transition temperature \cite{suppl}. Altogether, the electric field does not switch the disordered state (Fig. \ref{fig_scheme}(b) left) to the ordered state (Fig. \ref{fig_scheme}(b) right) but does suppress spin fluctuation in the ordered state. This interpretation is consistent with the results that the MECE peaks at a lower temperature than the ferroelectric phase boundary and that much smaller MECE is observed in the paraelectric phase (Fig. \ref{fig_phasediagram}(b)). Besides, the peak of MECE shifts to a lower temperature along the PE-FE phase boundary as the applied magnetic field increases, demonstrating that a magnetic field expands the operating temperature of MECE. By applying the optimum magnetic field for the refrigerant temperature, we can exploit the cooling effect for a wider range of temperature. The expansive operating temperature is desired for an effective Ericsson refrigeration cycle \cite{doi:10.1063/1.3654157}.


%
\begin{table*}[hbt!]
\caption{Operation parameters and  efficiencies $\eta$ for refrigeration effects in several materials: MECE, magnetocaloric effect (MCE), and adiabatic nuclear demagnetization (AND).
$T_{\mathrm{0}}$, operating temperature; $\Delta E$, change of electric field; $\Delta \mu _{\mathrm{0}}H$, change of magnetic field; $Q$, removed heat per unit volume. In AND, $Q$ is estimated by the equation: $Q = \int_{T_i}^{T_f}C_p(T)dT$, where $T_i$ ($T_f$) is the initial (final) temperature in an adiabatic demagnetization measurement \cite{ono1980}.
}
\begin{ruledtabular}
\begin{tabular}{llrrrrr}
\textrm{Refrigeration mechanisms}&
\textrm{Materials [Ref.]}&
\textrm{$T_{\mathrm{0}}$ (K)}&
\textrm{$\Delta E$ (kV/cm)}&
\textrm{$\Delta \mu _{\mathrm{0}}H$ (T)}&
\textrm{$Q$ ($\mathrm{J/cm^3}$)}&
\textrm{$\eta$ (\%)}\\
\colrule
MECE & $\mathrm{GdFeO_3}$[This work] & 2 & 25.6 &  & 0.012 & 1270\\
MCE & $\mathrm{GdFeO_3}$ \cite{PhysRevB.96.174405} & 5 &  & 2 & 0.28 & 14\\
MCE & $\mathrm{DyMnO_3}$ \cite{PhysRevB.84.235127} & 5 &  & 2 & 0.03 & 1.2\\
MCE & $\mathrm{GdVO_4}$ \cite{C6TC05182K} & 5 &  & 2 & 0.25 & 11\\
MCE & $\mathrm{Gd}$ \cite{PhysRevB.57.3478} & 294 &  & 2 & 12.7 & 580\\
AND & $\mathrm{PrCu_6}$ \cite{ono1980, Andres1972} & 0.02 &  & 5.5 & $1.9 \times 10^{-4}$ & $1.6 \times 10^{-3}$\\
\end{tabular}
\end{ruledtabular}
\label{tab:comparison}
\end{table*}

Table \ref{tab:comparison} compares MECE in $\mathrm{GdFeO_3}$ with MCE and adiabatic nuclear demagnetization (AND) of several rare-earth compounds. For evaluation of the different refrigeration mechanisms, we calculate energy-efficiency $\eta = |Q/W|$ \cite{Moya2015}, where $Q$ is removed heat per unit volume and $W$ is work done to apply or remove the electric/magnetic field. $W = V\int_{}{} EdD$ or $W = V\int_{}{} HdB$, where $V$ is the unit volume of the sample. $Q$ value of MECE is obtained by the direct measurement, while those of MCE and AND are estimated based on the previous literature reporting the temperature dependence of the heat capacity in a magnetic field and zero field. MECE shows a larger $\eta$ than MCE and AND do despite smaller $Q$. One must note that materials showing small $Q$s sometimes show higher efficiency \cite{Moya2015}.  Nevertheless, the present result implies that MECE is potentially applicable to energy-efficient cooling devices. The electric-field-driven entropy change may enable a higher-frequency refrigeration cycle than the magnetic-field driven cases, possibly compensating for small $Q$. In addition, an electric field can be applied locally to refrigerants. This may offer an advantage for local cooling.

In conclusion, we successfully demonstrated the MECE of multiferroic $\mathrm{GdFeO_3}$ in the experimental direct method. MECE shows a peak just below the ferroelectric transition temperature where an electric field suppresses the fluctuation of Gd moments. A magnetic field modulates the temperature at which MECE is maximized. The comparison between MECE, MCE, and AND reveals a high energy efficiency of MECE. Our research provides new insights to investigate new low energy-cost refrigeration techniques by using multicaloric effects.

T. K. was financially supported by MEXT Leading Initiative for Excellent Young Researchers (JPMXS0320200135), JSPS KAKENHI Grant-in-Aid for Young Scientists B (No. 21K13874). This work was partly supported by JSPS KAKENHI Grant-in-Aid for Scientific Research on Innovative Areas ”Quantum Liquid Crystals” (No. JP19H05826) and No. 19H01835. Measurements of X-ray Laue photographs, $P$, MECE were performed utilizing facilities of the Institute for Solid State Physics, the University of Tokyo. The measurement of heat capacity was carried out at the Cryogenic Research Center, the University of Tokyo.


\bibliography{ref_prl}
\ifarXiv
    

\section*{Supplementary material}

# Supplemental material for
# Magnetoelectrocaloric effect of multiferroic GdFeO$_3$

R. Ikeda, T. Kurumaji, Y. Tokunaga, and T. Arima

March 2, 2023

# Contents

# 1 MECE induced by electric fields in opposite directions

We show the sample temperature evolution when electric field $E = 16$ kV/cm was applied in positive and negative directions along the $c$ axis in Fig. S1. System is electrically unpolarized before the measurement. The measurements are conducted with a sample different from the one presented in the main text. The temperature changes were identical for both $E$ directions.

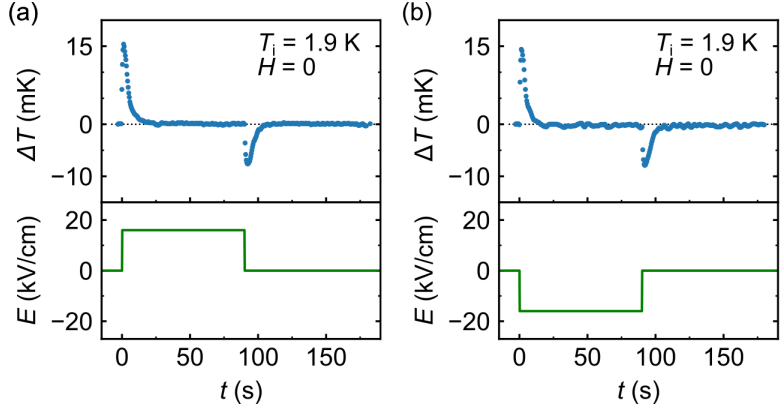

Fig. S1: Change of temperature ($\Delta T$) of the sample monitored with the resistance thermometer when the applied electric field ($E$) is in the positive direction along the $c$ axis (a) and in the negative direction (b). The bottom panels represent the applied electric field as a function of time.

# 2 MECE starting from the electrically polarized initial state and from an unpolarized initial state

Figure S2 shows the temporal evolution of the sample temperature change $\Delta T$ when the MECE measurement starts from unpolarized state (top panel) and polarized state (middle panel). The unpolarized state is obtained by cooling from the temperature above $T_\mathrm{N}^\mathrm{Gd}$ with $E = 0$. The poling field is $E = 16.8$ kV/cm. We observe a larger temperature increase for the unpolarized initial state than for the polarized initial state when the electric field is applied.

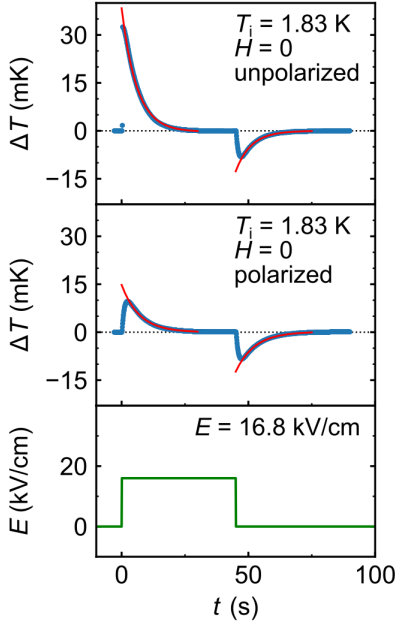

Fig. S2: Change of the sample temperature ($\Delta T$) and applied electric field ($E$) as a function of time ($t$) at initial sample temperature $T_\mathrm{i} = 1.85$ K in zero field. The initial state in the top (middle) panel is A' (A) in Fig. 3(a).

## 3 MECE in zero magnetic field for various initial temperatures

Figures S3 and S4 show MECE measurement data for various initial sample temperature $T_\mathrm{i}$. The sample is initially in the unpolarized state in Fig. S3 and in the polarized state with $E = 25.6$ kV/cm in Fig. S4. Red lines are fitted functions for estimating of $\Delta T_\mathrm{E}$ and $\Delta S$.

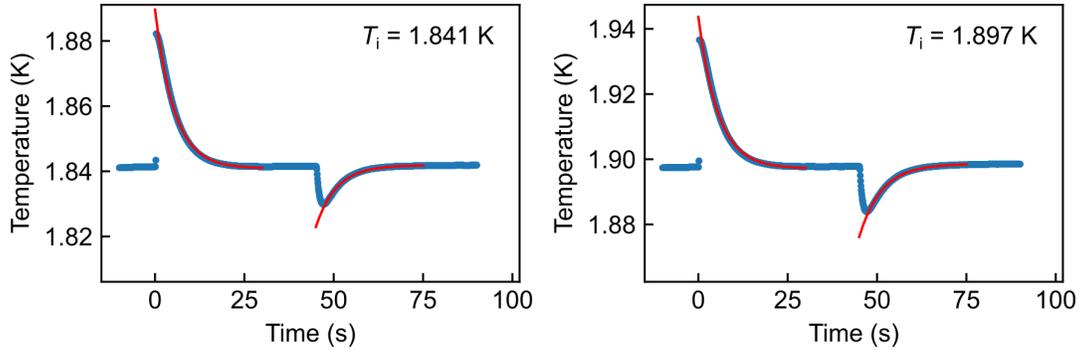

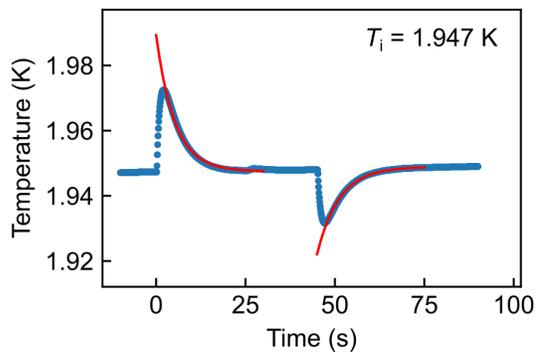
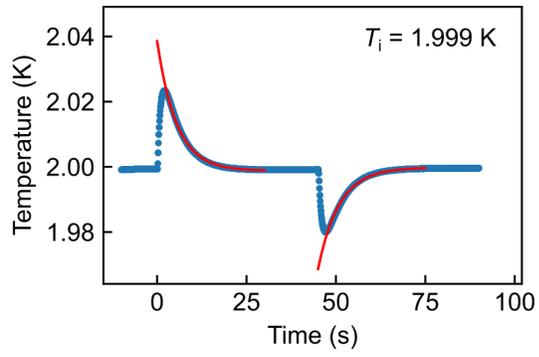
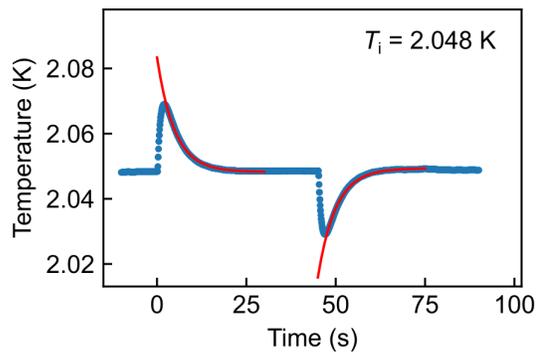
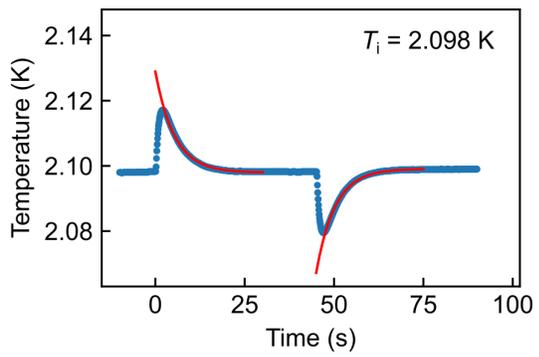
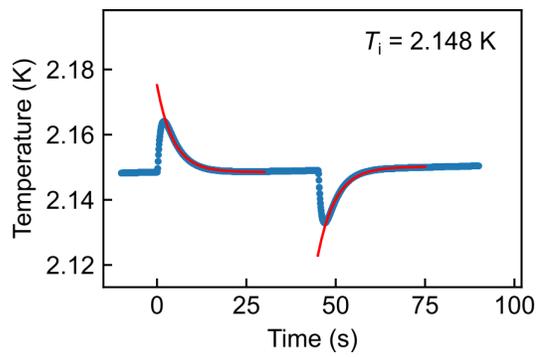
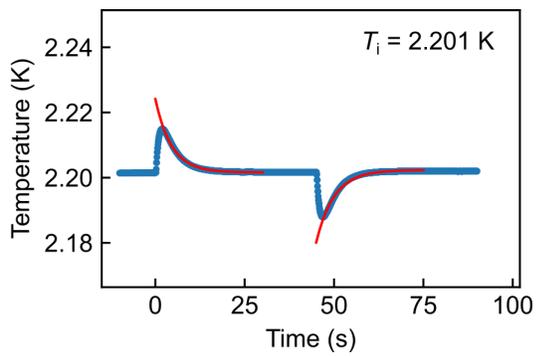
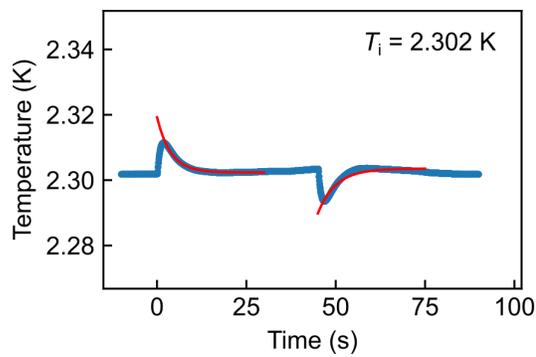
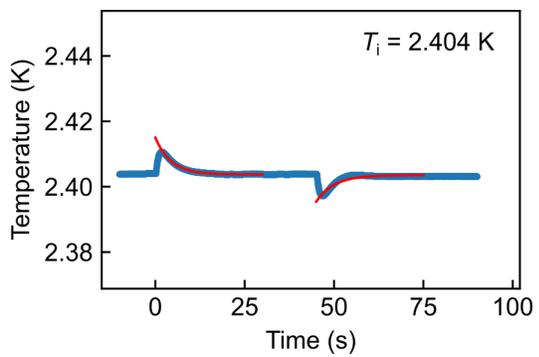

4

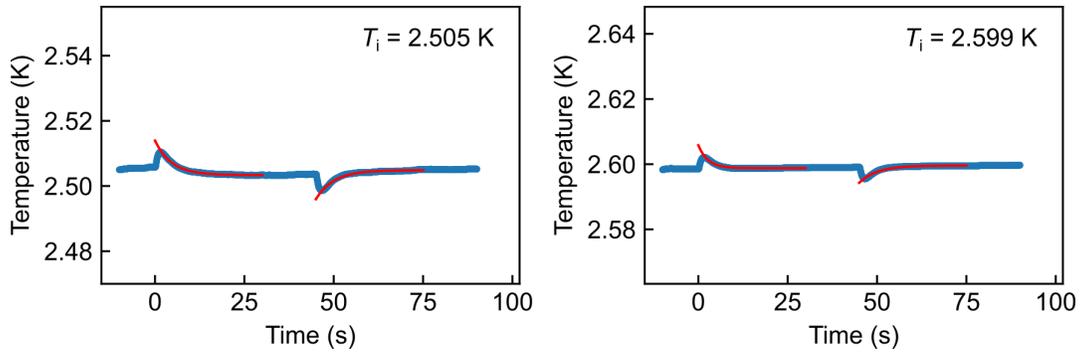

Fig. S3: Sample temperature as a function of time for different initial temperatures $T_\mathrm{i}$ in zero magnetic field. The sample is cooled from the temperature above the antiferromagnetic transition temperature of Gd moments at $E = 0$. The fitted functions were represented by red curves. The estimated $\Delta S$ values are shown by red circles and blue crosses in Fig. 4(a).

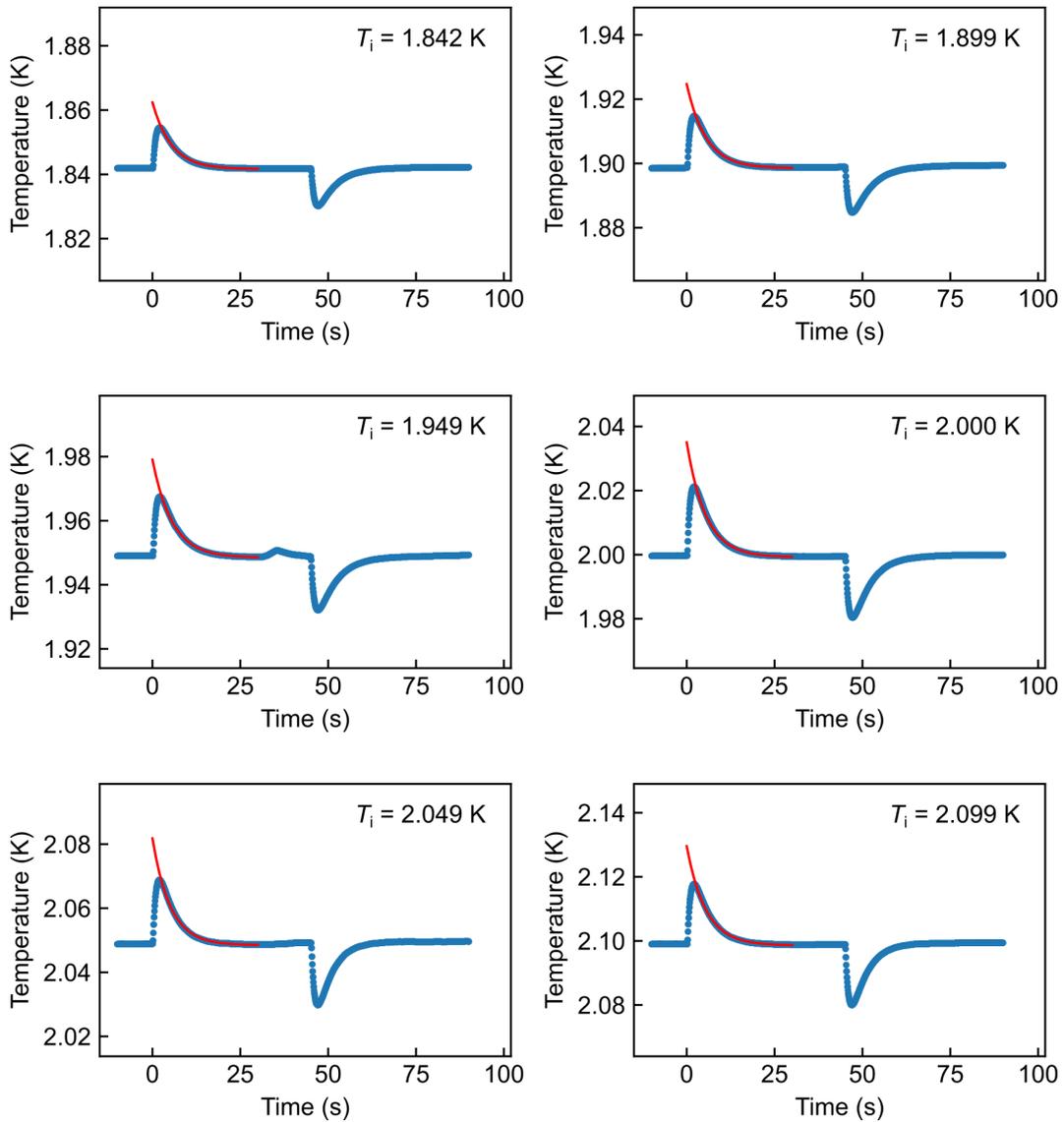

5

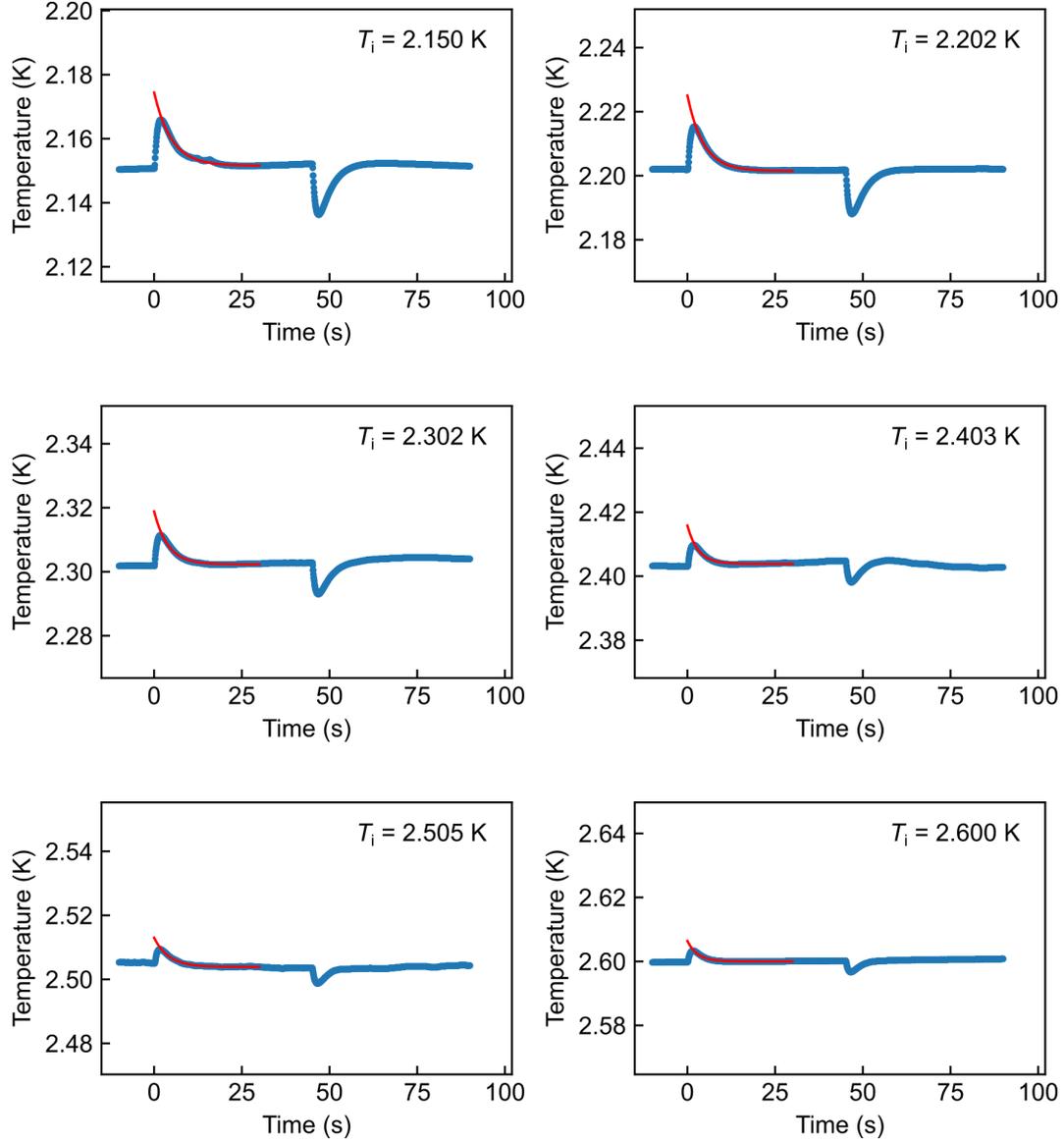

Fig. S4: Sample temperature as a function of time for different initial temperatures $T_\mathrm{i}$ in zero magnetic field. The sample is first polarized with $E = 25.6$ kV/cm at $T_\mathrm{i}$, and then the electric field is removed 60 seconds before the measurement. The fitted functions were represented by red curves. The estimated $\Delta S$ values are shown by light red open circles in Fig. 4(a).

## 4  MECE in various magnetic fields

We measure the time evolution of the sample temperature change $\Delta T$ in the MECE experiment in various magnetic fields, as shown in Fig. S5. The system is electrically unpolarized before the measurement. The applied electric field is 16.8 kV/cm. The larger temperature change is observed in a lower magnetic field, and negligibly small change is observed in 2 T.

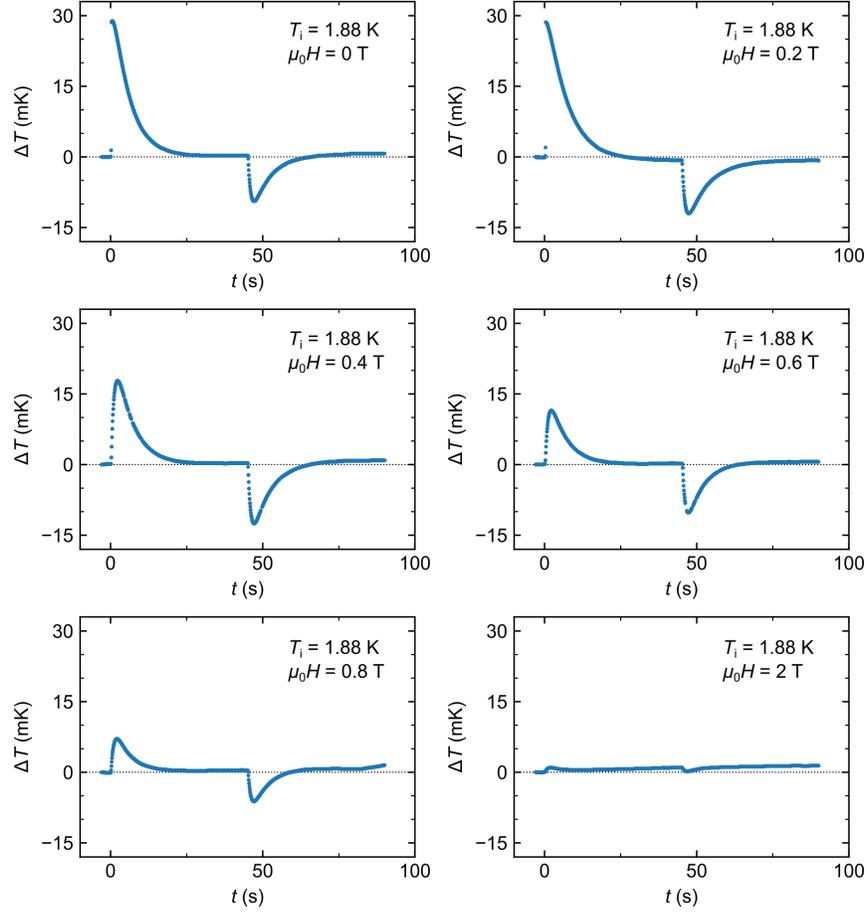

Fig. S5: Change of the sample temperature $\Delta T$ as a function of time $t$ in various magnetic fields $H$ parallel to the $a$ axis.

## 5  Heat capacity of the thermometer

The heat capacity of the thermometer (CX-1030-BR-HT) attached on the sample is shown in Fig. S6. In the temperature range of the MECE experiment, the heat capacity of the thermometer is negligibly small.

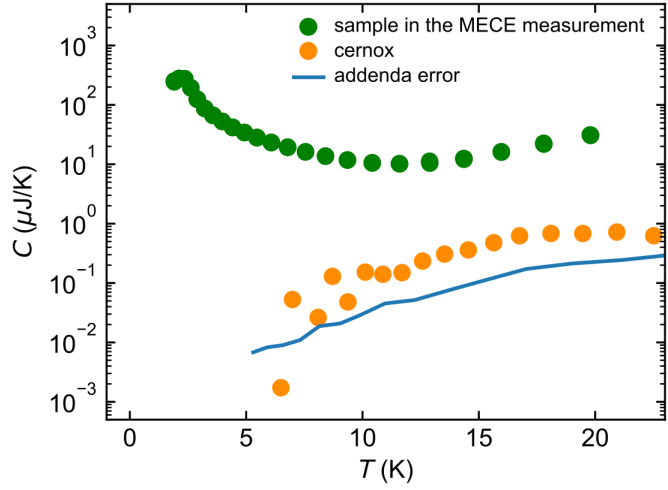

Fig. S6: Temperature dependence of heat capacity of the cernox thermometer in the MECE measurement. The heat capacity of the sample and the addenda error in the measurement of heat capacity of the thermometer are also shown for comparison.

# 6 Isothermal $P$-$E$ loops at several temperatures

Figure S7 represents $P$-$E$ hysteresis loops at several temperatures. The hysteresis curves are open at 1.8 K and 2 K, which indicates that the sample is in the ferroelectric states. At 2.2 K, just below the ferroelectric transition temperature, the hysteresis is closed but remains nonlinear. The polarization shows a linear response to the electric field at higher temperatures.

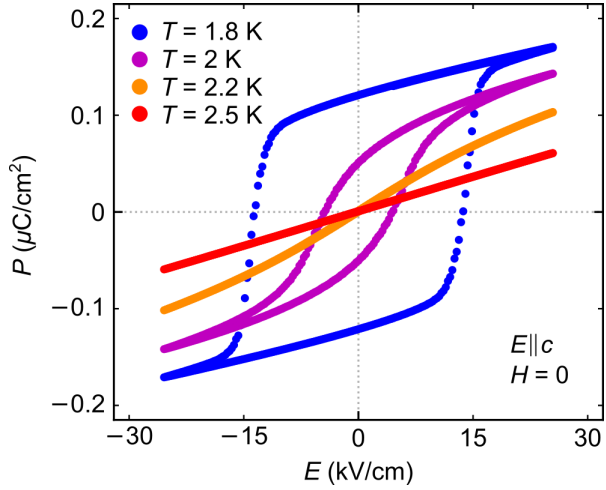

Fig. S7: $P$-$E$ hysteresis loops at various temperatures in zero magnetic field. The external electric field is applied parallel to the $c$ axis.

# 7 Work done by the electric field in the MECE measurement

We show the work done by the electric field in the $E$-application process of the MECE measurement in Fig. S8. We estimate the volume-normalized work $W$ as

$$W = \int E dP, \qquad (1)$$

where $E$ is the applied electric field and $P$ is the electric polarization. The polarization of the electrically polarized initial state is oriented parallel to the electric field, which is applied for the MECE measurement. The zero-$E$-field cooled state hosts negligibly small electric polarization. The work done on the zero-$E$-field cooled state agrees to that done on the electrically poled state above 2.2 K. Below 2.1 K, the discrepancy between the two cases become larger since zero-$E$-field cooling (electric poling) realizes electrically multidomain (single domain) state below the transition temperature of Gd moments. This is consistent with the $P$-$E$ hysteresis curves (Fig. S7). Extra work is necessary to rotate domains of the polarization antiparallel to $E$.

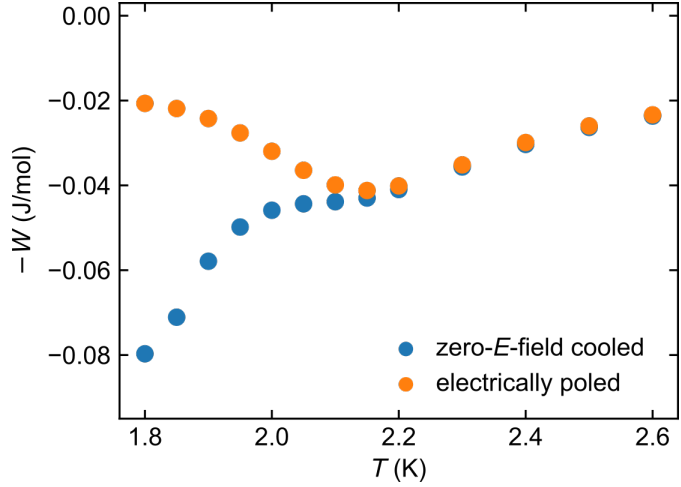

Fig. S8: Temperature dependence of the work done on the sample by the applied electric field. Orange (blue) circles represent that the initial state is obtained by an $E$-field poling with $E = 25.6$ kV/cm (zero-$E$-field cooling).

# 8 Temperature dependence of magnetization in electric fields

Temperature dependence of magnetization in electric fields is shown in Fig. S9. Below $T_\mathrm{N}^\mathrm{Gd}$, magnetization slightly decreases as a larger electric field is applied along the $c$ axis. Above $T_\mathrm{N}^\mathrm{Gd}$, magnetization are independent of the electric fields. The result suggests that an electric field enhances antiferromagnetic ordering of Gd moments only below the transition temperature.

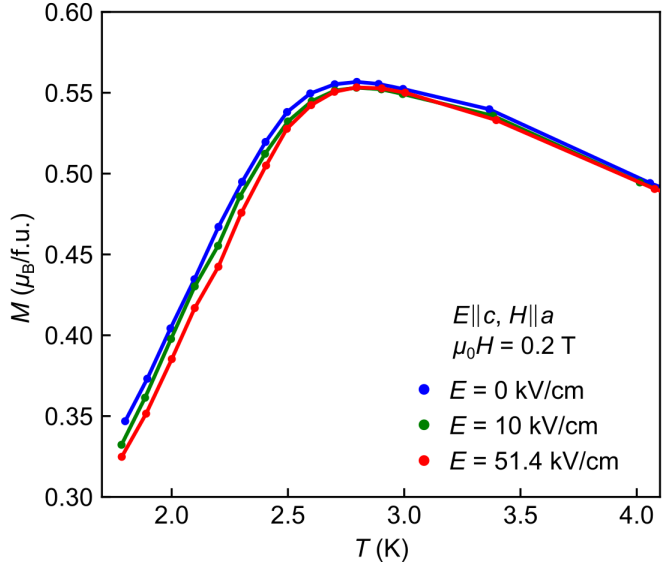

Fig. S9: Temperature dependence of magnetization with $\mu_0 H = 0.2$ T parallel to the $a$ axis in various electric fields applied along the $c$ axis.

# 9 Determination of the electric phase boundary

We determined the electric phase boundary (Fig. 4(b)) based on the peaks of the temperature dependence of heat capacity of $GdFeO_3$ measured at magnetic fields. Temperature dependence of specific heat of $GdFeO_3$ in $H$ parallel to the $a$ axis is shown in Fig. S10. The specific heat is measured on a sample of 1.2 mg in PPMS using the heat capacity option.

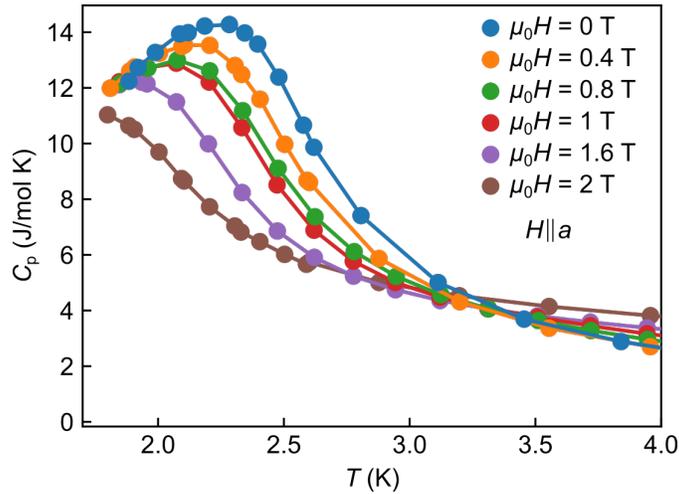

Fig. S10: Temperature dependence of the specific heat of $GdFeO_3$ in magnetic field parallel to the $a$ axis.

# 10 MECE in different domain states in terms of Fe moments

We compare MECE in a monodomain state with respect to Fe antiferromgnetic order and multidomain state. As the phase of antiferromagnetic arrangement of Fe moments is coupled to the weak

ferromagnetic moment direction, the phase is aligned by an external magnetic field of 9 T prior to the measurement. The results of the MECE measurement at 0 T for the two cases are shown in Fig. S11(a). Temperature changes are estimated as explained in the main text. We confirm identical temperature changes for both the states.

According to the previous discussion on the interaction energy [1], four combinations of signs of order parameters $P$, $M$ and $M'$ are allowed, as shown in Fig. S11(b). $P$ ($M$, $M'$) are the order parameter of exchange striction (Fe spins, Gd spins). In a multidomain state of $M$, all the domains I-IV are possible. When we apply an electric field, domain III (IV) changes to domain II (I). The transition from domain III to domain I (from domain IV to domain II) is almost impossible due to the faster reversal dynamics of Gd spins than that of Fe spins [1, 2]. If the ferromagnetic moment $M$ is positively aligned by an external magnetic field ($M+$), only domains I and IV are allowed before $E$-application. An applied electric field drives domain IV to domain I. The identical temperature change in multidomain and monodomain state of Fe antiferromagnetic phase suggests that the electric field driven motion of the composite domain wall of order parameters $P$ and $M'$ contributes to MECE.

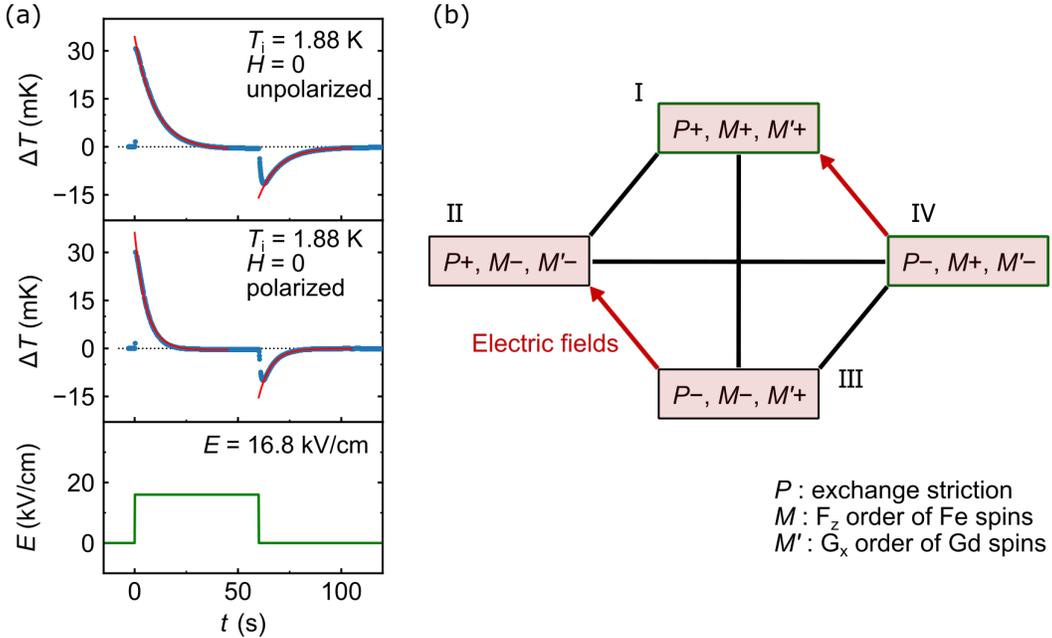

Fig. S11: (a) Change of temperature $\Delta T$ of the sample in the condition where weak ferromagnetic moments is unpolarized (upper) and magnetically polarized with the application of a magnetic field of 9 T (middle). The sample is initially in electrically unpolarized state. (b) Schematic diagram of domain states and their dynamics in GdFeO$_3$. $P$ ($M$, $M'$) denotes the order parameter of ferroelectricity (weak ferromagnetism of Fe spins, antiferromagnetic order of Gd spins), accompanying the sign of its order parameter. The transitions from domain III to domain II and from domain IV to domain I (red arrows), which are corresponding to simultaneous changes in $P$ and $M'$, are induced by $E$. Only the domains I and IV are allowed in a strong positive magnetic field.

## 11 MECE in the magnetic field of spin-flop transition

Figure S12 shows magnetic field dependence of the largest entropy change $\Delta S_{\text{max}}$ among the data obtained for various temperatures. The spin-flop transition, where both Gd and Fe spins rotate by 90 degrees, occurs at 0.5 T. We do not observe any anomalies related to the spin-flop transition.

$\Delta S_{\text{max}}$ shows a rapid decline above 1.2 T since the temperature where $\Delta S$ is the largest in each magnetic field is likely lower than the temperature range measured in the present study.

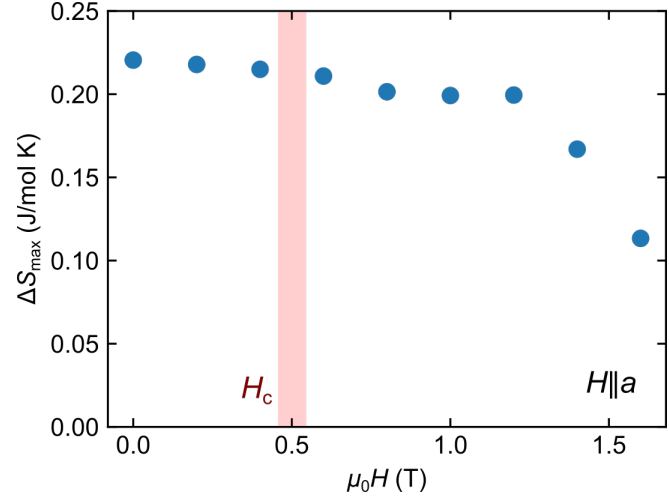

Fig. S12: Magnetic field dependence of the largest entropy change $\Delta S_{\text{max}}$ induced by MECE at each magnetic field, which is represented with orange circles in Fig. 4(b) of the main text. The magnetic field is applied parallel to the $a$ axis. $H_{\text{c}}$ denotes the magnetic field in which the spin-flop transition occurs.


\fi
\end{document}